\documentclass[12pt]{iopart}

\usepackage{epsfig}

\begin{document}

\title{Revisiting the one-dimensional diffusive contact process}

\author{W.G. Dantas and M.J. de Oliveira}

\address{Instituto de F\'{\i}sica,
Universidade de S\~{a}o Paulo,
Caixa Postal 66318
05315-970 S\~{a}o Paulo, S\~{a}o Paulo, Brazil}

\author{J.F. Stilck}
\address{Instituto de F\'{\i}sica, Universidade Federal Fluminense,
24210-340, Niter\'oi, RJ, Brazil}

\begin{abstract}
In this work we study the one-dimensional contact process with diffusion
using two different approaches to research the critical properties of
this model: the supercritical series expansions
and finite-size exact solutions. With special emphasis we look to the
multicritical point and its crossover exponent that characterizes
the passage between DP and mean-field critical properties.
This crossover occurs in the limit of infinite diffusion rate and our results
pointed $\phi=4$ as the better estimate for the crossover exponent in agreement
with computational simulations.  
\end{abstract} 


\maketitle

\section{Introduction}
In the last years there has been a growing interest in nonequilibrium
phase transitions \cite{md99,priv97}. 
The absence of a general theory
nonequilibrium systems originates a great number of 
open problems, even in one-dimensional systems 
that, in the equilibrium regime, generally are exactly solvable.
Usually, numerical simulations are a useful
technique in the study of phase transitions and critical
phenomena, and its power has been growing with the increasing capacity
of the computers and the development of new simulational
techniques. Non-equilibrium models are particularly suited for
simulations. However, others approaches may be complementary
in the study of these phenomena. Thus, it is interesting study 
the systems by other techniques. Among these alternative methods
we can cite the series expansions \cite{dj91}, which in some cases
lead to very precise estimates for the critical properties that characterize 
these transitions \cite{dj91,dickjenss} and the exact diagonalization of the
time evolution operator in finite-size systems \cite{carlon}.

A special case of the non-equilibrium models are the ones that exhibit 
absorbing states \cite{md99}, that is, states that may be reached in their 
dynamics, but transitions leaving them are forbidden. Since such
models do not obey detailed balance, they are intrinsically out of
equilibrium. 
The most studied system  for this class of problems is the so called 
contact process (CP) model \cite{h74}, a {\em toy model} for the 
spreading of an epidemic. This model displays a transition 
between an absorbing and an active state with critical exponents
belonging to the directed percolation (DP)
universality class \cite{dpjensen}. In addition, the CP 
model is related to the Schl\"ogl's lattice model for autocatalytic 
reactions \cite{s72} and the Reggeon Field Theory \cite{gt}.      

Many variants of this model have been studied \cite{dicktom91,mariofiore,
wgdstilckcross,yoon05}, most of them 
belonging to DP class also.  In fact, the 
robustness of this universality class is an evident characteristic of these
models. Such robustness is explained by the conjecture that 
all models with phase transitions between active and absorbing states 
with a scalar order parameter, short range interactions and no 
conservation laws belong to this class \cite{j81}.

One of these variants is the CP with diffusion \cite{dickjensdif},
which exhibits a critical line instead of a critical point. 
This line begins at the critical point of the model without diffusion 
and ends in the infinite diffusion rate limit, 
where the critical properties of the system  approaches those 
predicted by the mean-field approximation. For finite values of the
diffusion rate, the critical behavior of this model is dominated by
the DP universality class. This is not surprising, since the original
dynamic includes an intrinsic diffusion process. On the other hand, the
mean-field behavior in the limit of infinite diffusion 
may be understood considering that, since diffusion processes are
dominating the evolution of the system in this limit, creation
processes are effectively determined by the mean densities, as is
supposed in the mean-field approximation. 
This change of behavior at the infinite diffusion rate limit, 
between the critical behavior of the DP class and the one predicted by 
the mean-field approximation, characterizes a crossover of the
critical properties in  
the neighborhood of a multicritical point.  As in the equilibrium case
we may then write any density variable, in the neighborhood of a
multicritical point, as the following scaling form \cite{fk77}:
\begin{eqnarray}
g(\alpha,D)=(\alpha_c-\alpha)^{\theta}F\left(
\frac{D_c-D}{|\alpha_c-\alpha|^{\phi}}\right),
\end{eqnarray}
where $\alpha$ is a transition rate for annihilation of particles,
$D$ is the diffusion rate, ranging between 0 and 1, $\theta$ is 
a critical exponent associated to the density variable $g$, 
corresponding to the value predicted by the mean-field approximation
and $\phi$ is the crossover exponent. The scaling function 
$F(z)$ is singular at a point $z=z_0$ of its argument, such that the
critical line, 
in the neighborhood of the multicritical point, corresponds to
\begin{eqnarray}
(D_c-D)=z_0(\alpha_c-\alpha)^{\phi}.
\end{eqnarray}

One of the first studies of this problem was performed by
Dickman and Jensen \cite{dickjensdif}, who considered the model using   
supercritical series in $\alpha$ with the diffusion rate $D$ taken 
as a fixed parameter. Therefore, in their calculation series
expansions are derived for fixed values of the diffusion rate $D$, and
the analysis of this series leads, among other information, to the phase
diagram of the model with diffusion. 
However, they found that the fluctuations of the estimates provided by 
d-log Pad\'e approximants increase as the diffusion rate grows, 
so that the critical curve was obtained only up to 
$D\approx 0.8$. Since the crossover exponent $\phi$ characterizes the
critical curve close to the infinite diffusion rate limit $D \to
1$ no precise estimate of the crossover exponent was
possible. The disappointing performance of the Pad\'e approximants as
the multicritical point is approached is not surprising, since it is
known tha one-variable series analysis techniques are not effective
close to such points \cite{fk77}.
More recently \cite{mariofiore}, the model was simulated in a 
conservative ensemble. These simulations display smaller  
fluctuations in the estimates, making it possible to obtain the critical 
line up to values close to the multicritical point, furnishing 
the value $\phi=4.03(3)$ for the crossover exponent.
This result is consistent with the lower bound $\phi\geq 1$
predicted by Katori in \cite{kato94}.

In the present work, we obtain estimates for the critical line, 
exponents $\beta$ and $z$ as well as the crossover exponent $\phi$
for the one-dimensional CP with diffusion.
For this task we use two approaches: a two-variable supercritical
series and exact solutions for finite-size one-dimensional lattices.  
The supercritical series is analyzed using a partial differential 
approximants (PDA´s) \cite{fk77,s90,stilcksalinas}. This technique 
seems to be more appropriate for the analysis of a  
two-variable series with a multicritical behavior, as is shown by the
results obtained for other models \cite{wgdstilckcross,adlersalman}.     
This paper is organized as follows. In section II we present 
the model and the mean-field results, in section III we show
the derivation of the supercritical series and in section IV 
the analysis of this series is presented. In section V we discuss
the finite-size exact solution and their results for diffusive CP
model. Finally, in section VI, the conclusions and final discussions 
of this work may be found.  

\section{Definition of the model and mean-field results}

\begin{figure}
\begin{center}
\vspace*{1.1cm}
\epsfig{file=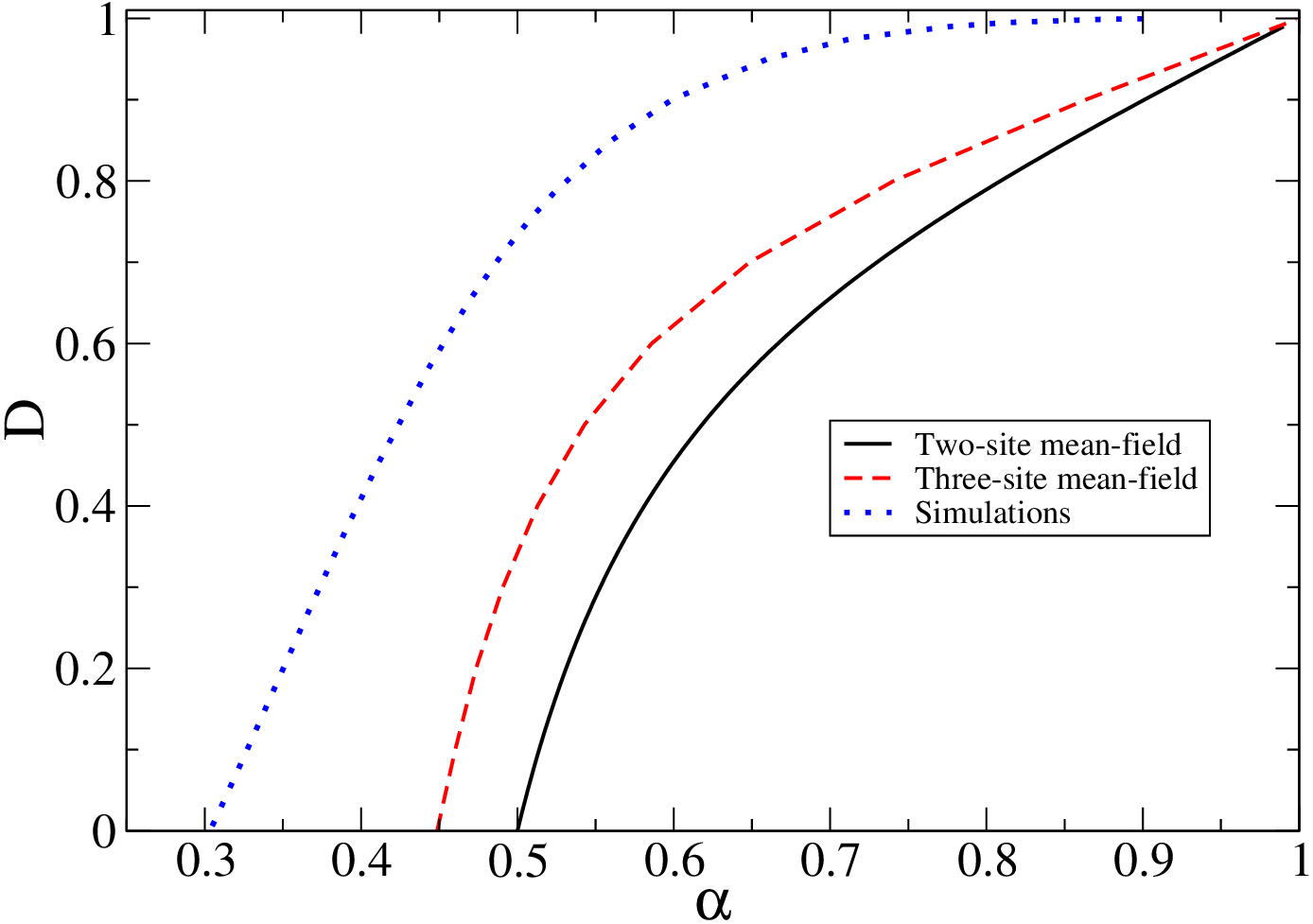,scale=0.35}\\
\vspace{1. cm}
\epsfig{file=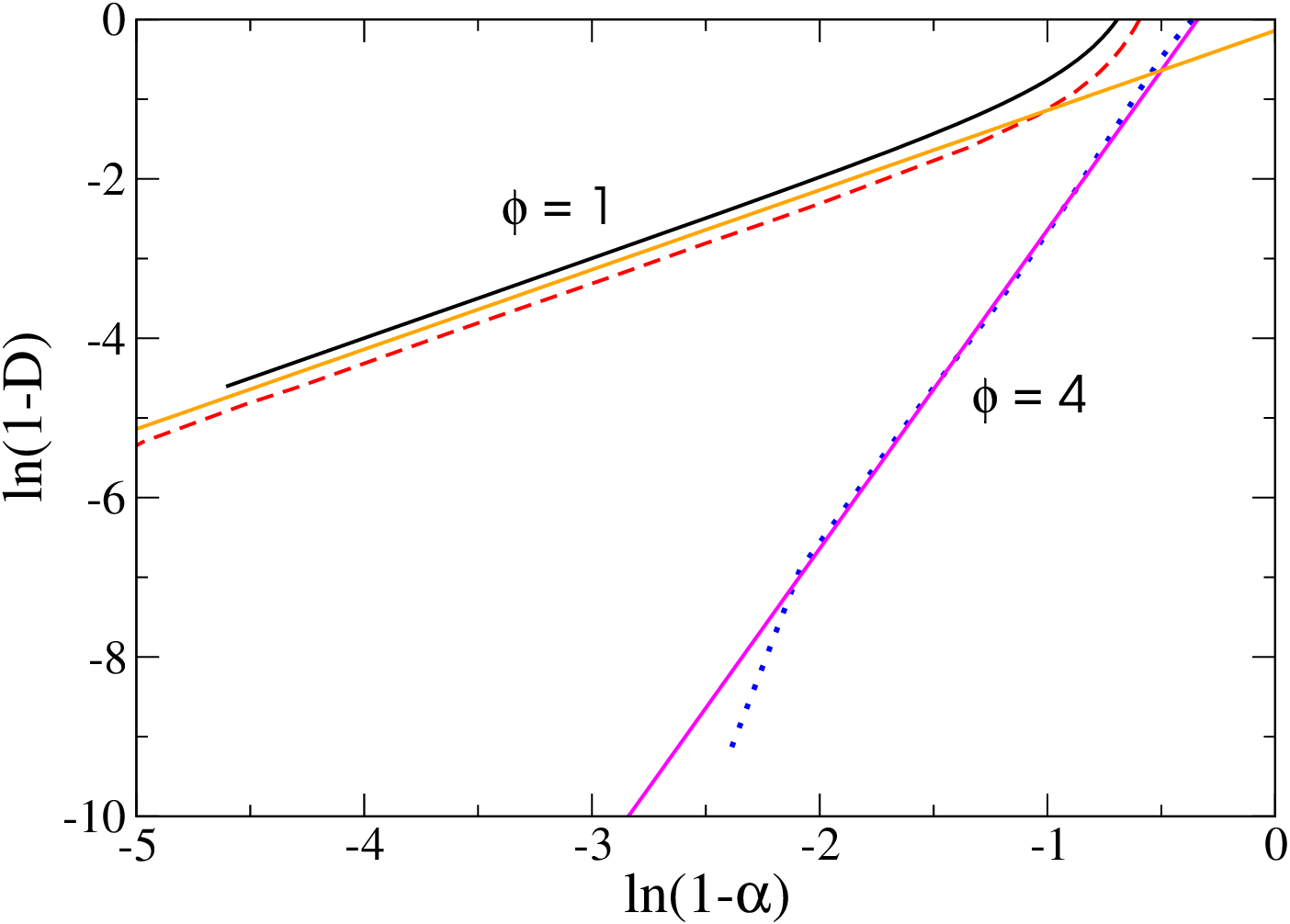,scale=0.35}
\vspace{0.1 cm}
\caption{Top: phase diagram obtained
using the two- and three-site mean-field approach 
and results of simulations \cite{mariofiore}. 
Bottom: the log-log plot of the same quantity 
is plotted and compared to the result $\phi=1$ and $\phi=4$
for these approaches.} 
\label{fig1}
\end{center}
\end{figure}

In a one-dimensional lattice each site can be empty or 
occupied by a particle, so that we will associate an occupation
variable $\eta_i=0,1$ to the site $i$.  
The evolution of the system is governed by
markovian local rules such that the particles are annihilated
with rate $\alpha$ and created in 
empty sites with a transition rate
$n/2$, where $n$ is the number of occupied nearest neighbors and $z$ 
is the total number of nearest neighbors.
In addition to these rules that define the CP, 
we include a diffusive process, allowing 
the hopping of particles to empty nearest neighbor sites 
at the rate $\tilde{D}=D/(1-D)$. The configuration such that all sites 
are empty is an absorbing state. The passage from an active
steady state, with a nonzero density of particles, to
the absorbing state defines a transition line in the $(\alpha,D)$
plane as shown in figure \ref{fig1}.

A mean-field approach for this model can be obtained at several
levels of approximation \cite{md99}.  In the one-site level
the role of the diffusion is irrelevant since 
it contributed equally to creation and annihilation of
particles at a given site $i$.  Already in the 
two-site level it is possible to determine the critical line 
by using as variables the parameters 
$\alpha$ and $D$. This line is described 
by the expression
\begin{eqnarray}
(1-D)=\frac{\alpha(1-\alpha)}{3\alpha-\alpha²-1}
\end{eqnarray}
and is very easy to show that in the neighborhood of 
the multicritical point, $(\alpha_c=1,D=1)$, the behavior of
this curve is given by the scaling relation
\begin{eqnarray}
(1-D)\sim(1-\alpha)^{\phi}
\end{eqnarray}
where $\phi=1$. On the three-site level an unitary crossover exponent
also appears, as  may be seen in figure \ref{fig1}.

This result is in accordance with the lower bound, $\phi\geq 1$,
determined by Katori \cite{kato94}. Comparing the mean-field approximation
with the simulational result \cite{mariofiore}, we observe that the first approach
always overestimates the supercritical region of the models.
Even in the higher diffusion region, the critical line
obtained using mean-field does not coincide with the simulational 
result. Actually, this coincidence occurs only in the multicritical
point.

In the next section we will derive the 
supercritical series in the variables $\lambda$ and $\tilde{D}$
to determine the value of this exponent and compare it
with that obtained in \cite{mariofiore,dickjensdif}. 

\section{Derivation of the supercritical series for the model}
\label{series}

We use the operator formalism proposed 
by Dickman and Jensen \cite{dickjenss} in order to
derive a supercritical series. To this end, we define the 
microscopic configuration of the lattice, $|\eta\rangle$, 
as the direct product of kets
$| \eta \rangle= \bigotimes_i | \eta_i \rangle$,
with the following orthonormality property,
$\langle \eta | \eta^\prime \rangle = \prod_i \delta_{\eta_i,\eta_i\prime}$.
The particle creation and annihilation operators at site
$i$ are defined as 
\begin{eqnarray}
A_i^\dagger |\eta_i \rangle &=&(1-\eta_i)|\eta_i+1\rangle,\nonumber \\
A_i |\eta_i\rangle &=& \eta_i|\eta_i-1\rangle.
\end{eqnarray}
In this formalism, the state of the system at time $t$ is
\begin{equation}
|\psi(t) \rangle = \sum_{\{\eta\}} p(\eta,t) |\eta \rangle,
\end{equation}
where $p(\eta,t)$ is the probability of a configuration $\eta$ at time $t$.
If we define the projection onto all possible states as
$\langle \; | \equiv \sum_{\{\eta\}} \langle\eta|$
then the normalization of the state of the system may be expressed as
$\langle \; |\psi \rangle =1$. In this notation, the master equation
for the evolution of the state is:
\begin{equation}
\frac{d |\psi(t)\rangle}{d t}=S|\psi(t)\rangle.
\label{me1}
\end{equation}
The evolution operator $S$ may be expressed in terms of the creation and
annihilation operators as $S=\mu R + V$ where
\begin{eqnarray}
\label{eqops}
R&=&\tilde{D}\sum_i (1-A_{i-1}^{\dagger}A_i)A_{i-1}A_i^{\dagger}+(1-A_{i+1}^{\dagger}A_i)
A_{i+1}A_i^{\dagger}+\nonumber\\
&+&\sum_i(A_i-A_i^{\dagger}A_i), \\
\label{eqopsv}
V &=& \sum_i (A_i^\dagger - A_iA_i^\dagger)(A_{i-1}^\dagger A_{i-1} +
A_{i+1}^\dagger A_{i+1}),
\end{eqnarray}
where $\mu\equiv 2\alpha$ and $\tilde{D}=D/\alpha$.

We notice that the operator $R$ diffuses $(01\to 10)$ or annihilates 
particles $(1\to 0)$, while the operator 
$V$ acts in the opposite way, generating particles $(0 \to 1)$. 
It is convenient to join the diffusion with the annihilation process 
to avoid ambiguities in the truncation of the series at a certain order.
For small values of the parameter $\mu$ the
creation of particles is favored, and the decomposition above is convenient
for a supercritical perturbation expansion. Using the equations 
(\ref{eqops}) and (\ref{eqopsv}) the action of each operator on a generical 
configuration $(\mathcal{C})$ is given by 
\begin{eqnarray}
R({\mathcal{C}})&=&\tilde{D}\left[\sum_i({\mathcal{C}}^{\prime}_i)+
\sum_j({\mathcal{C}}^{\prime r}_j)+
\sum_k({\mathcal{C}}^{\prime l}_k)\right]+\nonumber\\
&+&\sum_t({\mathcal{C}}^{\prime\prime}_t)
-[(r_1+2r_2)+\tilde{D}+r]({\mathcal{C}}),
\label{as0}
\end{eqnarray}
where the first sum is over $r_1$ sites with particles 
and one empty neighbor, the two next sums are over $r_2$ sites with 
particles and two empty neighbors and the last sum is over
all sites occupied by a particle.  Configuration $({\mathcal{C}}^{\prime}_i)$ 
is obtained moving the particle at the site $i$ to the single empty
neighbor site,
$({\mathcal{C}}^{\prime (r,l)}_i)$ is a configuration where the particle
at the site $i$ moved to the empty neighbor at the right ($r$) or at
the left ($l$) is replaced by a hole and one of the empty neighbors (at the
right or left) is occupied.
On the other hand, the action of operator $V$ is
\begin{equation}
V({\mathcal{C}})=\sum_i ({\mathcal{C}}^{\prime \prime\prime}_i)+2 \sum_j
({\mathcal{C}}^{\prime \prime\prime}_j) -(q_1+2q_2)({\mathcal{C}}),
\label{av}
\end{equation}
where the first sum is over the $q_1$ empty sites with one occupied
neighbor, the second sum is over the $q_2$ empty sites with two occupied
neighbors. Configuration $({\mathcal{C}}_i^{\prime \prime\prime})$ is obtained
occupying the site $i$ in configuration $(\mathcal{C})$.

To obtain a supercritical expansion for the ultimate survival probability of 
particles, we start by remembering that in order to access the long-time
behavior of a quantity, it is useful to consider its Laplace transform,

\begin{equation}
|\tilde{\psi}(s) \rangle = \int_0^\infty e^{-st} |\psi(t)\rangle.
\label{lt}
\end{equation}
Inserting the formal solution $|\psi(t)\rangle =e^{St} |\psi(0)\rangle$ of
the master equation (\ref{me1}) we find
\begin{equation}
|\tilde{\psi}(s) \rangle = (s-S)^{-1} |\psi(0)\rangle.
\label{tpsi}
\end{equation}
The stationary state $|\psi(\infty) \rangle \equiv \lim_{t \to \infty} |\psi(t)
\rangle$ may then be found by means of the relation 
\begin{equation}
|\psi(\infty) \rangle = \lim_{s \to 0} s |\tilde{\psi}(s) \rangle.
\end{equation}
A perturbative expansion may be obtained by assuming that
$|\tilde{\psi}(s) \rangle$  
can be expressed in powers of $\mu$ and using (\ref{tpsi}),
\begin{equation}
|\tilde{\psi}(s) \rangle = |\tilde{\psi}_0 \rangle+\mu |\tilde{\psi}_1 
\rangle +\mu^2 |\tilde{\psi}_2 \rangle + \cdots = (s- V -\mu R)^{-1}
|\psi(0) \rangle.  
\end{equation}
Since
\begin{equation}
(s- V -\mu R)^{-1}= (s-V)^{-1} \left[ 1 + \mu (s-V)^{-1}R 
+ \mu^2 (s-V)^{-2} R^2 + \cdots \right],
\end{equation}
we arrive at
\begin{equation}
|\tilde{\psi}_0 \rangle = (s-V)^{-1} |\psi(0)\rangle
\end{equation}
and
\begin{equation}
|\tilde{\psi}_n \rangle = (s-V)^{-1} R |\tilde{\psi}_{n-1} \rangle,
\end{equation}
for $n\geq 1$.
The action of the operator $(s-V)^{-1}$ on an arbitrary configuration
$({\mathcal{C}})$ may be found by noticing that
\begin{equation}
(s-V)^{-1} ({\mathcal{C}})=s^{-1}\left\{({\mathcal{C}})+ (s-V)^{-1}V
({\mathcal{C}})\right\}, 
\end{equation}
and using the expression \ref{av} for the action of the operator $V$, we get
\begin{equation}
(s-V)^{-1} ({\mathcal{C}})= s_q \left\{({\mathcal{C}}) + (s-V)^{-1} \left[
\sum_i 
({\mathcal{C}}^{\prime\prime\prime}_i)+2 \sum_j
({\mathcal{C}}^{\prime\prime\prime}_j) 
\right] \right\},
\label{sv}
\end{equation}
where the first sum is over the $q_1$ empty sites and one occupied
neighbor, the second sum is over the $q_2$ empty sites and two occupied
neighbors, and we define $s_q \equiv 1/(s+q_1+2 q_2)$.

It is convenient to adopt as the initial configuration a translational
invariant one with a single particle (periodic boundary conditions are
used). Now, looking at the recursive expression (\ref{sv}), we may notice 
that the operator $(s-V)^{-1}$ acting on any configuration generates 
an infinite set of configurations, and thus we are unable to calculate 
$|\tilde{\psi}\rangle$ in a closed form. However, it is possible calculate the 
extinction probability $\tilde{p}(s)$, which corresponds to the coefficient of 
the vacuum state $|0\rangle$. As happens also for  models
\cite{dickjenss,dj91}   
related to the CP, configurations with more than $j$ particles 
only contribute at orders higher than $j$, and since we are interested
in the ultimate survival 
probability for particles $P_\infty=1-\lim_{s \to 0} s \tilde{p}(s)$, $s_q$
may be replaced by $1/q$ in equation (\ref{sv}). An illustration of
this procedure may be found in a previous calculation \cite{wgdstilckcross}.

The algebraic operations described above is performed by a simple algorithm
which allow us to calculate 24 terms with a processing time to about
2 hours. Actually, the limiting factor in this operation is the memory 
required. In this way we define the coefficients $b_{i,j}$ for the ultimate
survival probability as:

\begin{eqnarray}
\label{pinf}
P_{\infty}=1-\frac{1}{2}\mu-\frac{1}{4}\mu^2-\sum_{i=3}^{24}\sum_{j=0}^{i-2}
b_{i,j}\mu^i\tilde{D}^j,
\end{eqnarray}
and they are listed in Table \ref{tab2}.

\begin{table}
\caption{Coefficients for the series expansion for ultimate survival
probability corresponding to the CP model with diffusion. The indexes
refer them to the equation (\ref{pinf}).}

\smallskip
\begin{tabular}{cccccc} 
\hline
\emph{i} & \emph{j} &$b_{i,j}$&\emph{i} & \emph{j} &$b_{i,j}$  \\
\hline
3 &0& 0.25000000000000000000$\times 10^0$      & &6&  0.10567643059624561630$\times 10^2$\\
  &1&-0.25000000000000000000$\times 10^0$      & &7&  0.34998474121093847700$\times 10^1$\\
4 &0& 0.28125000000000000000$\times 10^0$      &  &8& 0.12568359375000000000$\times 10^2$\\
  &1&-0.37500000000000000000$\times 10^0$      &11&0& 0.24775957118666096513$\times 10^1$\\
  &2& 0.37500000000000000000$\times 10^0$      &  &1&-0.71703082845177412707$\times 10^1$\\
5 &0& 0.34375000000000000000$\times 10^0$      &  &2& 0.12429764028158224676$\times 10^2$\\
  &1&-0.50781250000000000000$\times 10^0$      &  &3&-0.16676337106809967281$\times 10^2$\\   
  &2& 0.57812500000000000000$\times 10^0$      &  &4& 0.19435605471721252968$\times 10^2$\\
  &3&-0.62500000000000000000$\times 10^0$      &  &5&-0.15230894658300556443$\times 10^2$\\  
6 &0& 0.44726562500000000000$\times 10^0$      &  &6& 0.13923689787279895924$\times 10^2$\\
  &1&-0.76220703125000000000$\times 10^0$      &  &7&-0.24910518081099844778$\times 10^2$\\
  &2& 0.85058593750000000000$\times 10^0$      &  &8&-0.13853664539478481643$\times 10^2$\\
  &3&-0.83984375000000000000$\times 10^0$      &  &9&-0.23740234375000000000$\times 10^2$\\
  &4& 0.10937500000000000000$\times 10^0$\\
7 &0& 0.60223388671874955591$\times 10^0$      &12&0& 0.36488812342264926869$\times 10^1$\\
  &1&-0.11734619140625004441$\times 10^1$      &  &1&-0.11443929729648042226$\times 10^2$\\
  &2& 0.15190429687499997780$\times 10^1$      &  &2& 0.21418735868689896762$\times 10^2$\\  
  &3&-0.14140624999999984457$\times 10^1$      &  &3&-0.29831307350977681381$\times 10^2$\\
  &4& 0.10878906249999982236$\times 10^1$      &  &4& 0.34272964785342651339$\times 10^2$\\ 
  &5&-0.19687500000000000000$\times 10^1$      &  &5&-0.40398672142671166796$\times 10^2$\\
8 &0& 0.83485031127929687500$\times 10^0$      &  &6& 0.27855684964608855125$\times 10^2$\\
  &1&-0.18110389709472716202$\times 10^1$      &  &7&-0.13595902518316915319$\times 10^2$\\ 
  &2& 0.25234603881835981909$\times 10^1$      &  &8& 0.60935236387946176251$\times 10^2$\\ 
  &3&-0.29291381835937464473$\times 10^1$      &  &9& 0.40270442479922508028$\times 10^2$\\
  &4& 0.24864501953125062172$\times 10^1$      &  &10&0.45106445312500000000$\times 10^2$\\
  &5&-0.10754394531250017764$\times 10^1$      &13&0& 0.54293656084851154020$\times 10^1$ \\ 
  &6& 0.36093750000000000000$\times 10^1$      &  &1&-0.18322144692814863021$\times 10^2$\\
9 &0& 0.11814667913648828623$\times 10^1$      &  &2& 0.36259195896082665911$\times 10^2$\\
  &1&-0.28569926950666579835$\times 10^1$      &  &3&-0.54866931326313050477$\times 10^2$\\
  &2& 0.42781094621729156557$\times 10^1$      &  &4& 0.67130818799164941879$\times 10^2$\\
  &3&-0.48761836864330092567$\times 10^1$      &  &5&-0.64293858561553619779$\times 10^2$\\
  &4& 0.54410674483687788694$\times 10^1$      &  &6& 0.81638245065997452343$\times 10^2$\\
  &5&-0.48496839735243062464$\times 10^1$      &  &7&-0.59206203158103356543$\times 10^2$\\
  &6& 0.11848958333333414750$\times 10^0$      &  &8&-0.11386872839957611347$\times 10^2$\\
  &7&-0.67031250000000000000$\times 10^1$      &  &9&-0.15017302299755911577$\times 10^3$\\
10&0& 0.16988672076919952847$\times 10^1$      &  &10&-0.10358591484729181786$\times 10^3$\\
  &1&-0.45030223008843064392$\times 10^1$      &  &11&-0.8611230468750000000$\times 10^2$\\
  &2& 0.73700355965291270977$\times 10^1$      &14&0&0.8132542219307161701600$\times 10^1$\\
  &3&-0.92486491500105252328$\times 10^1$      &  &1&-0.29467694610727896531$\times 10^2$\\
  &4& 0.87502182305104447835$\times 10^1$      &  &2& 0.62075441392908530247$\times 10^2$\\
  &5&-0.93882905701060082038$\times 10^1$      &  &3&-0.96520364146752442025$\times 10^2$\\
\end{tabular}
\label{tab2}
\end{table}

\begin{table}
\smallskip
\begin{tabular}{cccccc} 
\hline
\emph{i} & \emph{j} &$b_{i,j}$&\emph{i} & \emph{j} &$b_{i,j}$  \\
\hline
14&4& 0.12740413805065173847$\times 10^3$       &  &3&-0.55737662171810006839$\times 10^3$\\
  &5&-0.14872799806680012580$\times 10^3$       &  &4& 0.80584299514354984240$\times 10^3$\\ 
  &6& 0.11055259565281477308$\times 10^3$       &  &5&-0.10663609887965685630$\times 10^4$\\ 
  &7&-0.15322889380857196784$\times 10^3$       &  &6& 0.12359653482806177180$\times 10^4$\\
  &8& 0.15125687744198603468$\times 10^3$       &  &7&-0.86967640846259655518$\times 10^3$\\
  &9& 0.12100775973170790678$\times 10^3$       &  &8& 0.16397080915531578285$\times 10^4$\\
  &10&0.36717258144235222517$\times 10^3$       &  &9&-0.72176857313912660175$\times 10^3$\\
  &11&0.24952145042880511028$\times 10^3$       &  &10&-0.39467789553376610456$\times 10^3$\\  
  &12&0.16504858398437500000$\times 10^3$       &  &11&-0.37241724510229601037$\times 10^4$\\
15&0& 0.12275012836144505002$\times 10^2$       &  &12&-0.43433579993156563432$\times 10^4$\\
  &1&-0.47363165128788978109$\times 10^2$       &  &13&-0.49111545112523144780$\times 10^4$\\
  &2& 0.10546586796137503939$\times 10^3$       &  &14&-0.28643360597728060384$\times 10^4$\\
  &3&-0.1762739398241103288$\times 10^3$        &  &15&-0.11834527587890625000$\times 10^4$\\
  &4& 0.23298609118631188153$\times 10^3$       &18&0& 0.435207828742268674200$\times 10^2$\\
  &5&-0.27071715385838172097$\times 10^3$       &  &1& -0.20009555228747112210$\times 10^3$\\
  &6& 0.33267266081610591755$\times 10^3$       &  &2& 0.51666085550451919062$\times 10^3$\\
  &7&-0.18014432368848466126$\times 10^3$       &  &3&-0.98221829678260564833$\times 10^3$\\
  &8& 0.24323790718884771422$\times 10^3$       &  &4& 0.15468813789798582548$\times 10^4$\\
  &9&-0.43351480900655008099$\times 10^3$       &  &5&-0.19452358996876919264$\times 10^4$\\
  &10&-0.48110863117407200207$\times 10^3$      &  &6& 0.23142581042234874076$\times 10^4$\\
  &11&-0.88513499744041246231$\times 10^3$      &  &7&-0.29269143532222028625$\times 10^4$\\
  &12&-0.57707323452079549497$\times 10^3$      &  &8& 0.11929249112512670763$\times 10^4$ \\
  &13&-0.31740112304687500000$\times 10^3$      &  &9&-0.33469260525748682085$\times 10^4$\\
16&0& 0.18620961415130427241$\times 10^2$       &  &10&0.22662397929431922421$\times 10^4$ \\
  &1&-0.76547748518027589171$\times 10^2$       &  &11&0.33534053058169365613$\times 10^4$\\
  &2& 0.17936794520034777634$\times 10^3$       &  &12&0.10497352179518215053$\times 10^5$\\
  &3&-0.30967915791812674797$\times 10^3$       &  &13&0.11603782689493993530$\times 10^5$\\
  &4& 0.45127497217248452444$\times 10^3$       &  &14&0.11326231062527707763$\times 10^5$ \\
  &5&-0.53698491310493750461$\times 10^3$       &  &15&0.62269615466431168898$\times 10^4$ \\
  &6& 0.51912309945306844838$\times 10^3$       &  &16&0.22929397201538085938$\times 10^4$ \\
  &7&-0.74776029427868388666$\times 10^3$       &19&0& 0.66930218067969633466$\times 10^2$ \\
  &8& 0.31390176074091152714$\times 10^3$       &  &1&-0.32354897975245813768$\times 10^3$\\
  &9&-0.23066668763466492464$\times 10^3$       &  &2& 0.88042629554806353553$\times 10^3$\\
  &10&0.12806582574012411442$\times 10^4$       &  &3&-0.17413898514485581472$\times 10^4$\\
  &11&0.15251186818533849419$\times 10^4$       &  &4& 0.27574673463423659996$\times 10^4$\\
  &12&0.21006816788719602300$\times 10^4$       &  &5&-0.39760107863198809355$\times 10^4$\\
  &13&0.12983815231244370807$\times 10^4$       &  &6& 0.45478180647223589403$\times 10^4$\\
  &14&0.61213073730468750000$\times 10^3$       &  &7&-0.44829538123020120111$\times 10^4$\\
17&0& 0.28405733950686048672$\times 10^2$       &  &8& 0.71439950475520599866$\times 10^4$\\
  &1&-0.12342415559750365617$\times 10^3$       &  &9&-0.11433220550468702186$\times 10^4$\\
  &2& 0.30541526863109334045$\times 10^3$       &  &10&0.58699584417021997069$\times 10^4$\\
\end{tabular}
\end{table}

\begin{table}
\smallskip
\begin{tabular}{cccccc} 
\hline
\emph{i} & \emph{j} &$b_{i,j}$&\emph{i} & \emph{j} &$b_{i,j}$  \\
\hline
19&11&-0.80007309950477119855$\times 10^4$        &  &15&-0.19550932821134978440$\times 10^6$\\
  &12&-0.13849586891129882133$\times 10^5$        &  &16&-0.17865808405461237999$\times 10^6$\\ 
  &13&-0.28630836746692133602$\times 10^5$        &  &17&-0.13001245697926016874$\times 10^6$\\ 
  &14&-0.29699369329023520550$\times 10^5$        &  &18&-0.60334798749708410469$\times 10^5$\\
  &15&-0.25808082841629722679$\times 10^5$        &  &19&-0.16853935146331787109$\times 10^5$\\
  &16&-0.13385541065918856475$\times 10^5$        &22&0&  0.24876519640902955643$\times 10^3$\\
  &17&-0.44510006332397460938$\times 10^4$        &  &1& -0.13849806980532957823$\times 10^4$\\
20&0&  0.10337399908883011790$\times 10^3$        &  &2&  0.42491525612070690840$\times 10^4$\\  
  &1& -0.52548922262251915072$\times 10^3$        &  &3& -0.95509490156905540061$\times 10^4$\\
  &2&  0.14837268484319015442$\times 10^4$        &  &4&  0.17072122607398174296$\times 10^5$\\
  &3& -0.30794482965317188246$\times 10^4$        &  &5& -0.25259991312968326383$\times 10^5$\\
  &4&  0.51825134723219762236$\times 10^4$        &  &6&  0.36959825267281092238$\times 10^5$\\
  &5& -0.70384028435684167562$\times 10^4$        &  &7& -0.39724946702482979163$\times 10^5$\\
  &6&  0.95392163017937873519$\times 10^4$        &  &8&  0.41082488201681495411$\times 10^5$\\
  &7& -0.10775090683332531626$\times 10^5$        &  &9& -0.69996295461125846487$\times 10^5$\\
  &8&  0.72306238600702890835$\times 10^4$        &  &10&-0.47433844045513687888$\times 10^4$ \\
  &9& -0.17556031650766155508$\times 10^5$        &  &11&-0.94472082453477501986$\times 10^5$\\
  &10& 0.45576565336850308086$\times 10^3$        &  &12& 0.18231631734823597071$\times 10^5$\\
  &11&-0.68853229355527937514$\times 10^4$        &  &13& 0.74507403911089320900$\times 10^5$ \\
  &12& 0.27725591978402942914$\times 10^5$        &  &14& 0.28127919612702319864$\times 10^6$\\
  &13& 0.46620453043310422800$\times 10^5$        &  &15& 0.40344708716105029453$\times 10^6$\\
  &14& 0.75787894589888033806$\times 10^5$        &  &16& 0.49346542769156675786$\times 10^6$ \\
  &15& 0.73699439769879478263$\times 10^5$        &  &17& 0.42523410507562680868$\times 10^6$\\
  &16& 0.58191226154708187096$\times 10^5$        &  &18& 0.28817317888963740552$\times 10^6$\\
  &17& 0.28519783989193914749$\times 10^5$        &  &19& 0.12690253004534545471$\times 10^6$ \\
  &18& 0.86547234535217285156$\times 10^4$        &  &20& 0.32865173535346984863$\times 10^5$\\
21&0&  0.15998830271612598608$\times 10^3$        &23&0&  0.38696067609131841891$\times 10^3$\\
  &1& -0.85206464313625656359$\times 10^3$        &  &1& -0.22521288931067451813$\times 10^4$ \\
  &2&  0.25270411365871177622$\times 10^4$        &  &2&  0.72239403631839231821$\times 10^4$  \\
  &3& -0.53883333002320678133$\times 10^4$        &  &3& -0.16540644823073282168$\times 10^5$  \\
  &4&  0.93069165541648508224$\times 10^4$        &  &4&  0.30965978110985601234$\times 10^5$ \\
  &5& -0.14256926805592749588$\times 10^5$        &  &5& -0.49715203648917136888$\times 10^5$\\
  &6&  0.16894039347875652311$\times 10^5$        &  &6&  0.63140021723358484451$\times 10^5$  \\
  &7& -0.21057424914324066776$\times 10^5$        &  &7& -0.90160716109902248718$\times 10^5$ \\
  &8&  0.2676311383571557235$\times 10^5$         &  &8&  0.96095356087432839558$\times 10^5$\\
  &9& -0.76173492536429366737$\times 10^4$        &  &9& -0.63690648044614848914$\times 10^5$\\
  &10& 0.42006937485571783327$\times 10^5$        &  &10& 0.18830794813510467065$\times 10^6$\\
  &11&-0.12698208721358355433$\times 10^4$        &  &11& 0.57163083968398816069$\times 10^5$\\
  &12&-0.53474107987358220271$\times 10^4$        &  &12& 0.19078924010583921336$\times 10^6$\\
  &13&-0.90835383726464555366$\times 10^5$        &  &13&-0.11462564856718564988$\times 10^6$\\
  &14&-0.14158471202173284837$\times 10^6$        &  &14&-0.34295621170633088332$\times 10^6$\\
\end{tabular}
\end{table}

\begin{table}
\smallskip
\begin{tabular}{cccccc} 
\hline
\emph{i} & \emph{j} &$b_{i,j}$&\mbox{ }&\mbox{ }&\mbox{ }\\
\hline
23&15&-0.83015840201854216866$\times 10^6$\\
  &16&-0.11002212525564364623$\times 10^7$\\
  &17&-0.12227985275158903096$\times 10^7$\\
  &18&-0.99733247054306510836$\times 10^6$\\
  &19&-0.63431234698974736966$\times 10^6$\\
  &20&-0.26563714369463571347$\times 10^6$\\
  &21&-0.64165338807344436646$\times 10^5$\\
24&0&  0.60509199550873199769$\times 10^3$\\ 
  &1& -0.36606232348859748527$\times 10^4$\\
  &2&  0.12139375556095224965$\times 10^5$\\
  &3& -0.29330923818458009919$\times 10^5$\\ 
  &4&  0.55510496259245075635$\times 10^5$\\
  &5& -0.89523794399456470273$\times 10^5$\\  
  &6&  0.13690056986421268084$\times 10^6$\\
  &7& -0.14950913041347730905$\times 10^6$\\
  &8&  0.20276027262469305424$\times 10^6$\\  
  &9& -0.24868958310934680048$\times 10^6$\\ 
  &10& 0.44658906685524416389$\times 10^5$\\ 
  &11&-0.50473239786133670714$\times 10^6$\\
  &12&-0.20583143731838543317$\times 10^6$\\
  &13&-0.31319042987920023734$\times 10^6$\\ 
  &14& 0.51903724071582528995$\times 10^6$\\
  &15& 0.12406518345377091318$\times 10^7$\\
  &16& 0.23558257104359627701$\times 10^7$\\  
  &17& 0.29049027840298512019$\times 10^7$\\  
  &18& 0.29832956761808455922$\times 10^7$\\
  &19& 0.23110049545479607768$\times 10^7$\\
  &20& 0.13877345160856433213$\times 10^7$\\
  &21& 0.55381177327087428421$\times 10^6$\\
  &22& 0.12541407130479812622$\times 10^6$\\
\hline
\end{tabular}
\end{table}

\section{Analysis of the series}
\label{analysis}

To obtain estimates of the critical properties, specially
the critical line, from the supercritical series 
for the ultimate survival probability as given by the 
equation (\ref{pinf}), we  initially use
d-log Pad\'e approximants. These approximants
are defined as ratios of two polynomials 

\begin{eqnarray}
F_{LM}(\lambda)=\frac{P_L(\lambda)}{Q_M(\lambda)}=
\frac{\sum_{i=0}^{L}p_i\lambda^i}  
{1+\sum_{j=1}^M q_j\lambda^j}=f(\lambda).
\end{eqnarray}
In our case the function $f(\lambda)$ represents the series for
$\frac{d}{d\lambda} 
\ln P_{\infty}(\lambda)$. As $f(\lambda)$ is a function of one variable, 
we fix the value of $\tilde{D}$ to calculate these approximants. For a fixed
value of $\tilde{D}$ one pole of the approximant $F$ will correspond to the
critical point while the associated residue will be the critical
exponent $\beta$.    
We calculate approximants with $L+M\leq 24$, restricting our
calculation to diagonal or close to diagonal approximants, which
usually display a better convergence. Thus
$L=M+\xi$, where $\xi=0,\pm 1$ and with $D=\tilde{D}/(1+\tilde{D})$
ranging between 0 and 0.8. Examples of
estimates for the critical values of $\mu$ obtained from these
approximants is given in Table \ref{tab11} for different values of 
the diffusion.

\begin{table}
\begin{center}
\begin{tabular}{cccc}
\hline
D&L&M&$\mu_c$\\
\hline
0  &10&10&0.60645\\
   &11&11&0.60646\\
0.1&10&10&0.62267\\
   &11&11&0.62266\\
0.7&10&10&0.85353\\
   &11&11&0.84513\\
0.8&10&10&0.96256\\
   &11&11&0.94246\\
\hline
\end{tabular}
\caption{Estimates for critical points obtained by d-log Pad\'e
  approximants. Note that  
  as the value of $D=\tilde{D}/(1+\tilde{D})$ grows the dispersion in the
  value estimates also increases.}
\label{tab11}
\end{center}
\end{table}

\begin{figure}[h!]
\begin{center}
\vspace*{0.9cm}
\includegraphics[scale=0.5]{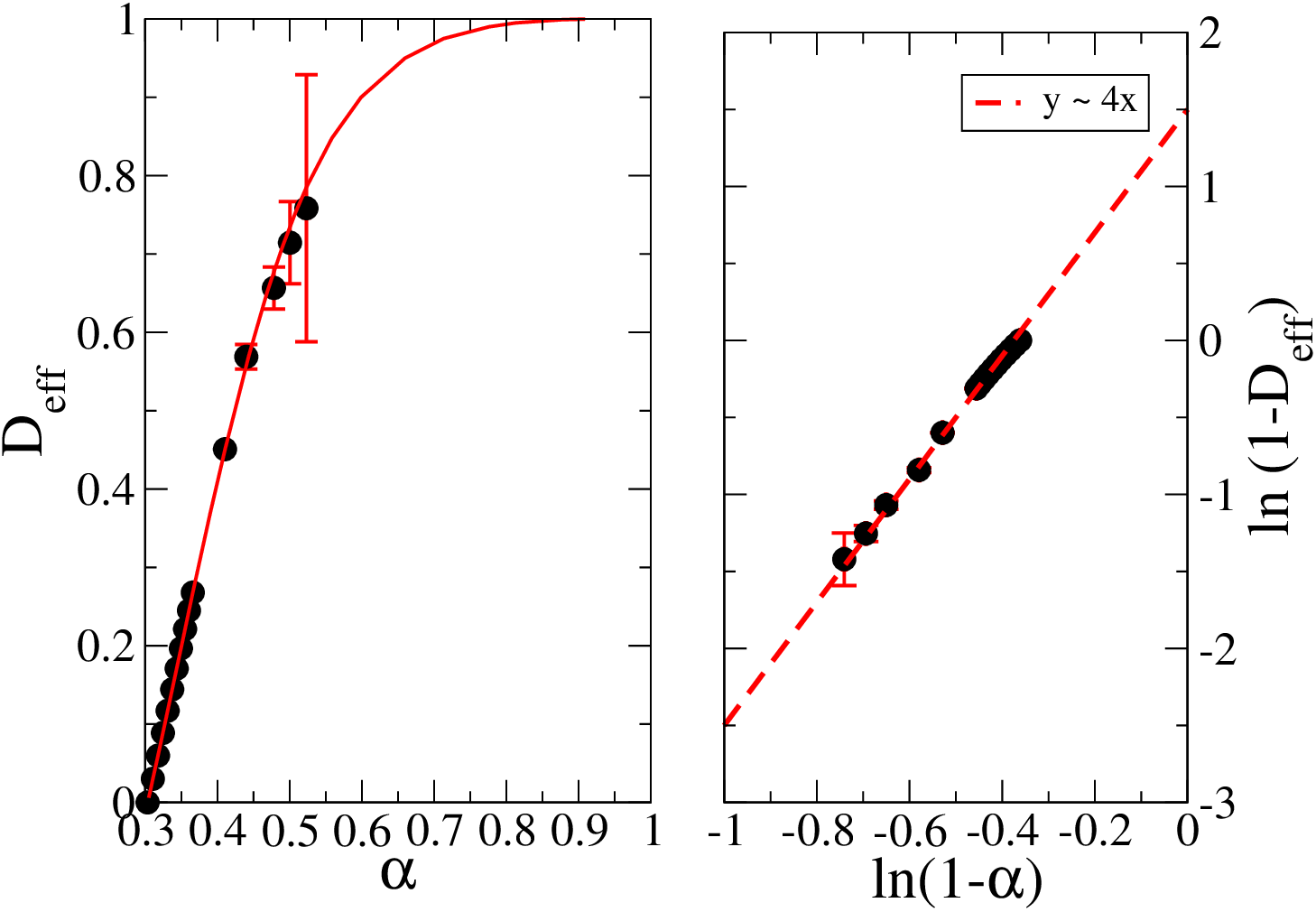}
\caption{At left panel we have the phase diagram obtained
by simulations (solid curve) and through Pades approximants for the supercritical
series (circles). At right the log-log plot of the
same quantity is plotted showing that value of the crossover exponent $\phi=4$
is a reasonable estimate for this data.}
\label{fig2}
\end{center}
\end{figure}

\begin{figure}[h!]
\begin{center}
\vspace*{0.9cm}
\includegraphics[scale=0.4]{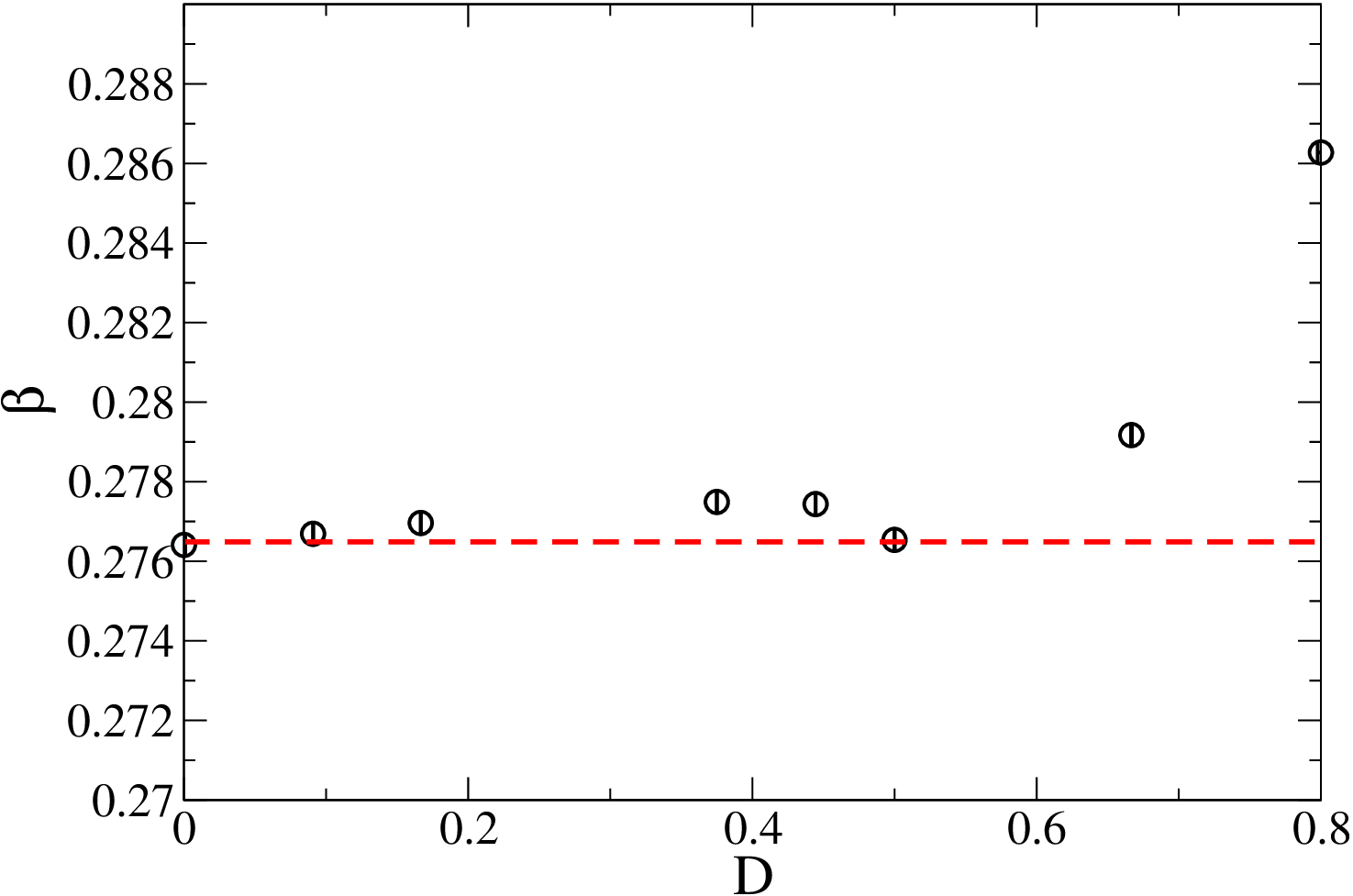}
\caption{Estimates of the exponent $\beta$ for different diffusion rates.
The dashed line indicates the DP value for this exponent.}
\label{beta}
\end{center}
\end{figure}

For each value of the diffusion rate, we calculate about eight 
approximants, obtaining the estimate of $\mu_c$ associated to
diffusion as an arithmetic average of results furnished by these set
of approximants and error bar associated to it corresponds to standard
deviation of the estimates.  
From this we obtain the phase diagram shown
in the figure \ref{fig2}. With the purpose of comparison 
with the results coming from the conservative simulations
\cite{mariofiore} 
we use the variables $\alpha\equiv\mu/2$ 
and $D_{eff}=\alpha\tilde{D}/(1+\alpha\tilde{D})$. We also
calculate the $\beta$ exponent associated to the order parameter
for differents values of $D$ exhibited in the figure \ref{beta}. 
Similar to what happens with the critical point a growing fluctuation
for high diffusion rate values is visible.  

Turning back to the discussion about the critical line, we see that
for higher values of the diffusion the dispersion increases, 
and estimates with larger error bars are found.  Nevertheless, the
exponent $\phi=4$ seems to describe well the calculated points of the
critical line.  However, this result
would be different if we used approximants to series closer to  
the infinite diffusion rate limit.  
This error in the approximants for high values of the 
diffusion rate is attributed to the
alternated sign of the series terms \cite{dickjensdif}.  Another 
explanation would come from the fact that in the
neighborhood of a multicritical point the reduction of a 
two-variable series to one variable leads to very poor
estimates of the critical properties \cite{wgdstilckcross} close to a
multicritical point.
To overcome this problem we analyze the series 
using Partial Differential Approximants (PDA's) \cite{fk77}, 
that generalize the d-log Pad\'e approximants for a two-variable 
series.  These approximants are defined by the following equation  
\begin{equation}
P_{\mathbf L}(x,y)F(x,y)=Q_{\mathbf M}(x,y)\frac{\partial F(x,y)}{\partial x} +
R_{\mathbf N}(x,y) \frac{\partial F(x,y)}{\partial y},
\label{pda}
\end{equation}
where $P$, $Q$, and $R$ are polynomials in the variables $x$ and $y$ with the
set of nonzero coefficients ${\mathbf L}$, ${\mathbf M}$, and ${\mathbf N}$,
respectively. The coefficients 
of the polynomials are obtained by substitution of the series expansion
of the quantity which is going to be analyzed 
\begin{equation}
f(x,y)=\sum_{k,k^\prime=0} f(k,k^\prime)x^k y^{k^\prime}
\end{equation}
into equation (\ref{pda}) and requiring the equality to hold for a
set of indexes defined as ${\mathbf K}$. This procedure leads to a system
of linear 
equations for the coefficients of the polynomials, and since the coefficients
$f_{k,k^\prime}$ of the series are known for a finite set of indexes
this places 
an upper limit to the number of coefficients in the polynomials. Since the
number of equations has to match the number of unknown coefficients,
the numbers of elements in each set must satisfy $K=L+M+N-1$ (one
coefficient is fixed arbitrarily). An additional issue, 
which is not present in the one-variable case, is the symmetry of the
polynomials. Two frequently used options are the triangular and the
rectangular arrays of coefficients. The choice of these symmetries may
be related to the symmetry of the series itself \cite{s90}. In our case,
the series presents a triangular symmetry when express in terms of the
variables $x=\alpha$ and $y=\alpha\tilde{D}/(1+\alpha\tilde{D})$.

It is possible to show \cite{s90} that we can  
determine the multicritical properties using the equation \ref{pda} 
and the hypothesis of that in the neighborhood of the multicritical point, 
the function $f$ obeys  the following scaling form
\begin{equation}
f(x,y) \approx |\Delta \widetilde{x}|^{-\nu}Z\left( \frac{|\Delta
\widetilde{y}|}{|\Delta \widetilde{x}|^\phi} \right),
\label{mcs}
\end{equation}
where
\begin{equation}
\Delta \widetilde{x}=(x-x_c )-(y-y_c)/e_2,
\end{equation}
and
\begin{equation}
\Delta \widetilde{y}=(y-y_c)-e_1(x-x_c).
\end{equation}
Here $\nu$ is the critical exponent of the quantity described by $f$ when
$\Delta \widetilde{y}=0$, $e_1$ and $e_2$ are the scaling slopes \cite{fk77}
and 
$\phi$ is the crossover exponent. 

\begin{figure}[h!]
\begin{center}
\vspace*{0.9cm}
\includegraphics[scale=0.5]{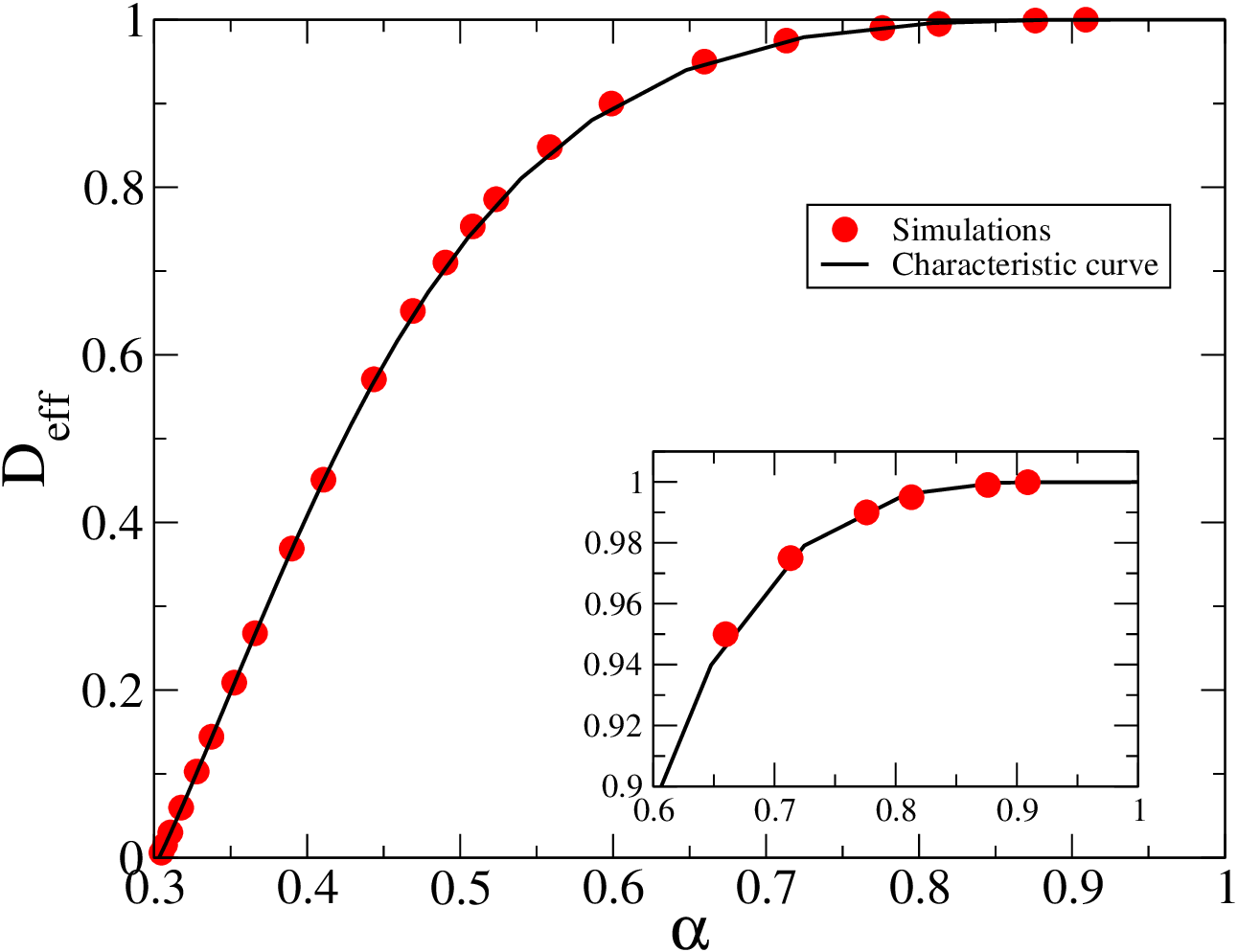}
\vspace*{0.2cm}
\caption{Comparison between a characteristic curve (solid line) and
a numerical simulation result (circles). The coincidence is evident,
including the region close to the infinite diffusion limit (inset).}
\label{fig3}
\end{center}
\end{figure}

On the other side, our calculation was successful when we use 
the method of the characteristics to integrate equation 
(\ref{pda}).  This is made by introducing a timelike variable $\tau$, 
so that a family of curves is obtained in the plane 
$(x(\tau),y(\tau))$ (the characteristics).  
Such curves obey to the equations
\begin{eqnarray}
\frac{dx}{d \tau} &=& Q_M(x(\tau),y(\tau)),\nonumber\\
\frac{dy}{d\tau}  &=& R_N(x(\tau),y(\tau)).
\label{char}
\end{eqnarray}
It is possible to show that integrating the equations (\ref{char}) from a 
specific point of the critical line, the resulting 
characteristic is equivalent to the the critical line. In figure
\ref{fig3} we show a comparison between a characteristic curve and the 
simulational result \cite{mariofiore}. 
The number of elements in the sets of the calculated approximants was varying
as follows: $K=55-190$, $M=N=20-53$ and $L=15-54$.

In each of these curves, 
we calculate his inclination in the neighborhood of the multicritical
point, determining a value for the exponent $\phi$ and the mean value
for this exponent results as $\phi=4.02\pm0.13$, consistent with 
simulations in the particle conserving ensemble
\cite{mariofiore}. 

\begin{figure}[h!]
\begin{center}
\vspace*{0.9cm}
\includegraphics[scale=0.4]{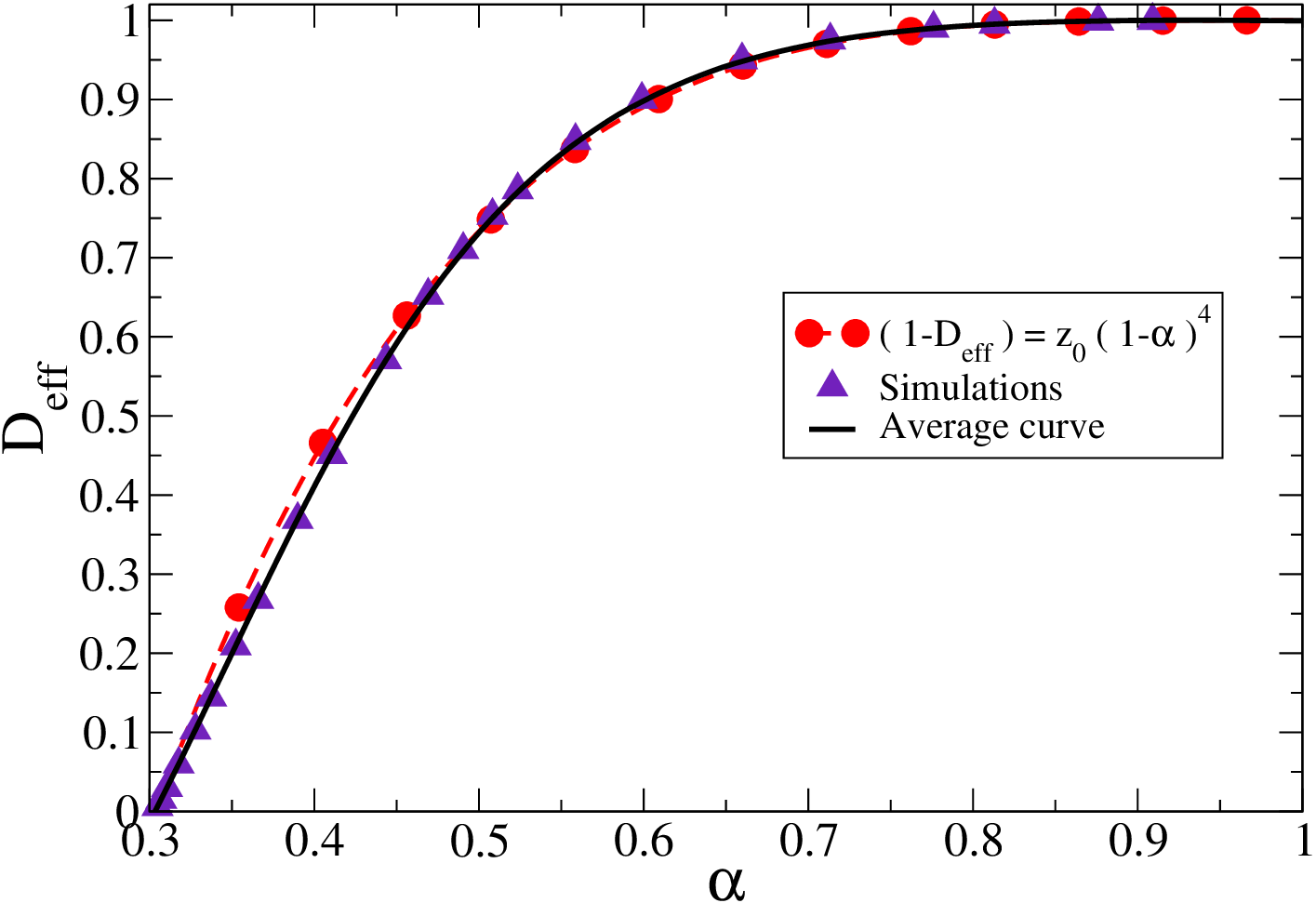}
\\
\vspace*{1.1cm}
\includegraphics[scale=0.4]{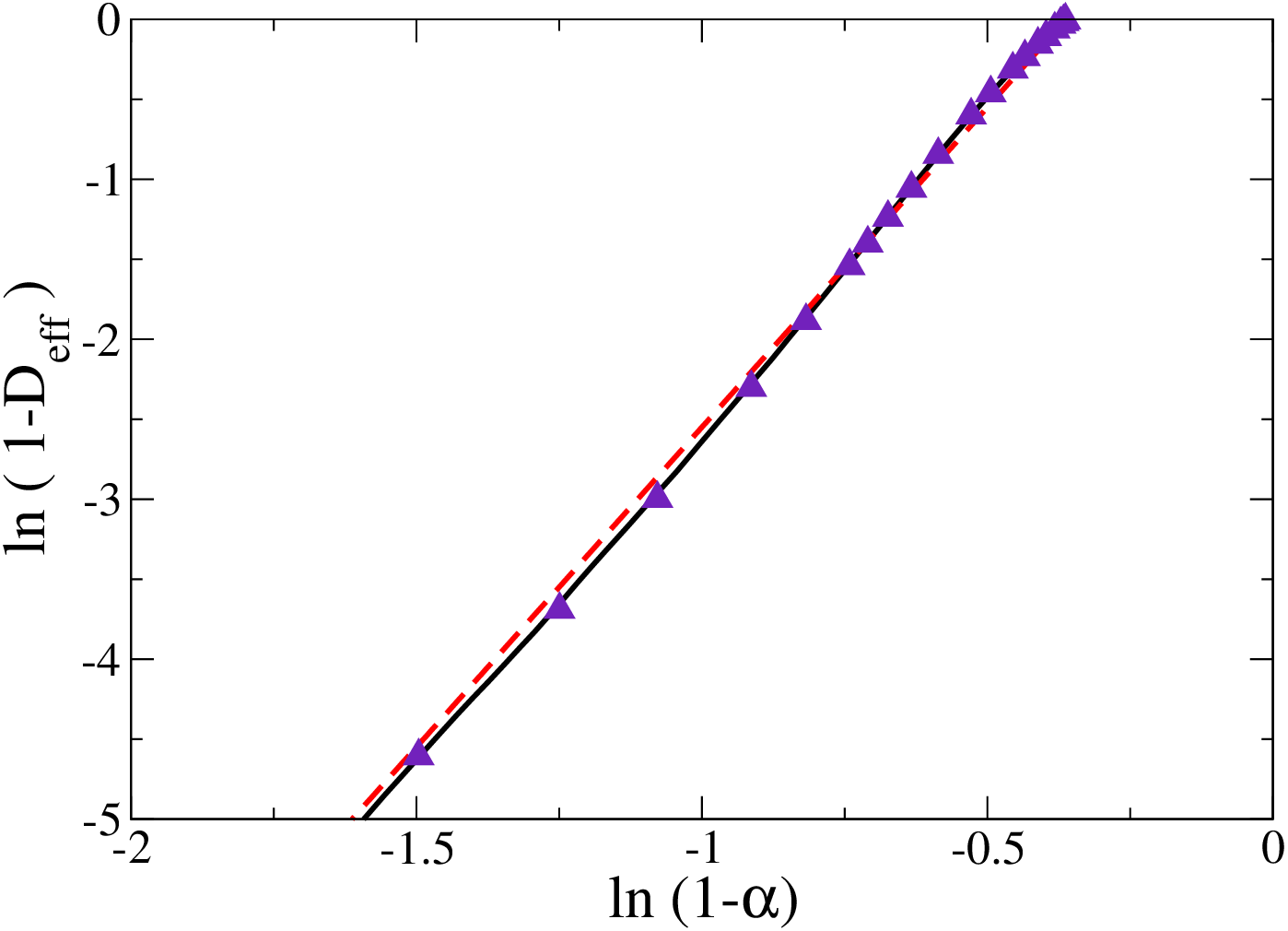}
\vspace*{0.2cm}
\caption{Comparison between a characteristic `average curve' (solid line), 
the numerical simulation results (triangles) and the curve obtained
from the scaling form $(1-D_{eff})\sim(1-\alpha)^4$(circles). 
At the right panel we see that $\phi=4$ seems to be a good estimative for
the crossover exponent.}
\label{fig4}
\end{center}
\end{figure}

Using all the characteristic curves calculated we derive an `average curve', 
calculating for each value of $\alpha$ on arithmetic average for $D_{eff}$. 
This curve is shown in the figure \ref{fig4} jointly with the result
originating  
from the simulation and with the scaling form
$(1-D_{eff})\sim(1-\alpha)^4$.  In the same 
figure we see that the exponent $\phi=4$ is well fitted to the results of 
simulation and of the PDA's. This scaling form is based on the argument of the 
scaling function $Z$ presented in the equation (\ref{mcs}), where $\phi=4$ and
$z_0$ is a parameter properly chosen. We remark that this scaling form 
coincides with the characteristic curve and with the simulation even in the
weak diffusion regime. This is somewhat surprising since its validity
would be expected only in the neighborhood of the multicritical point.

Unfortunately, even using the algorithm proposed by Styer \cite{s90} 
we were not able to obtain precise estimates for 
the crossover exponent $\phi$ from the scaling form shown in 
equation (\ref{mcs}). However, integrating a set of
approximants, we could determine the characteristic curves   
whose initial point is coincident with the critical point of the CP 
without diffusion of particles. These curves are estimates 
for the critical line of the CP model with diffusion going 
beyond the values achieved in \cite{dickjensdif} 
and \cite{mariofiore} and corroborating the initial result of this last 
reference in that $\phi\approx 4$.

\section{Exact solution for finite systems}

An interesting approach to study stochastic systems
is the exact solution of models with increasing numbers
of sites L followed by extrapolations to the thermodynamic
limit \cite{carlon}. This is accomplished by considering operator $S$
and its eigenvalues. For process governed by the master equation
vanishing eigenvalues,$\mu_0 = 0$, of the evolution operator correspond 
to stationary states of the system. For finite
systems with absorbing states, only these states are stationary,
and no active stationary state is found.
To study the transition between an active and a stationary
state, we may consider the behavior of the eigenvalue with the
second smallest absolute value $\Gamma=|\mu_1|$ of the operator. This
eigenvalue is related to the quasi-stationary state \cite{vidigal} and will
eventually become degenerate with $\mu_0$ in the thermodynamic
limit, originating the phase transition.

In a one-dimensional lattice with $L$ sites, we construct a set
of configurations that works as basis of the operator. Using periodic
boundary conditions some of these configurations will be related
for a symmetry $C_L$ over cyclic rotations. These symmetries
reduces the number of the independent vectors of the basis. 

\begin{figure}[h!]
\begin{center}
\vspace*{0.9cm}
\includegraphics[scale=0.4]{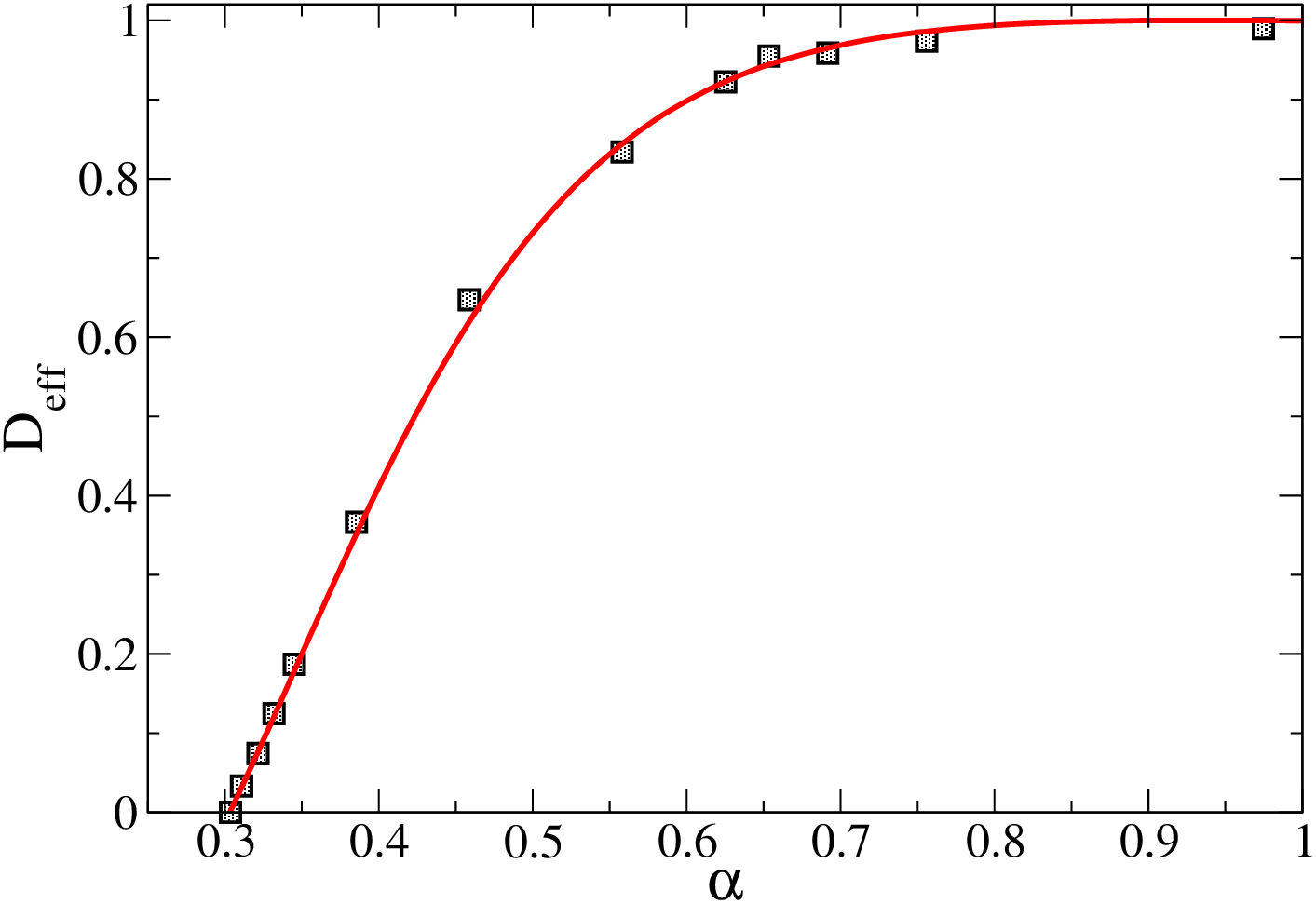}\\
\vspace*{1.15cm}
\includegraphics[scale=0.4]{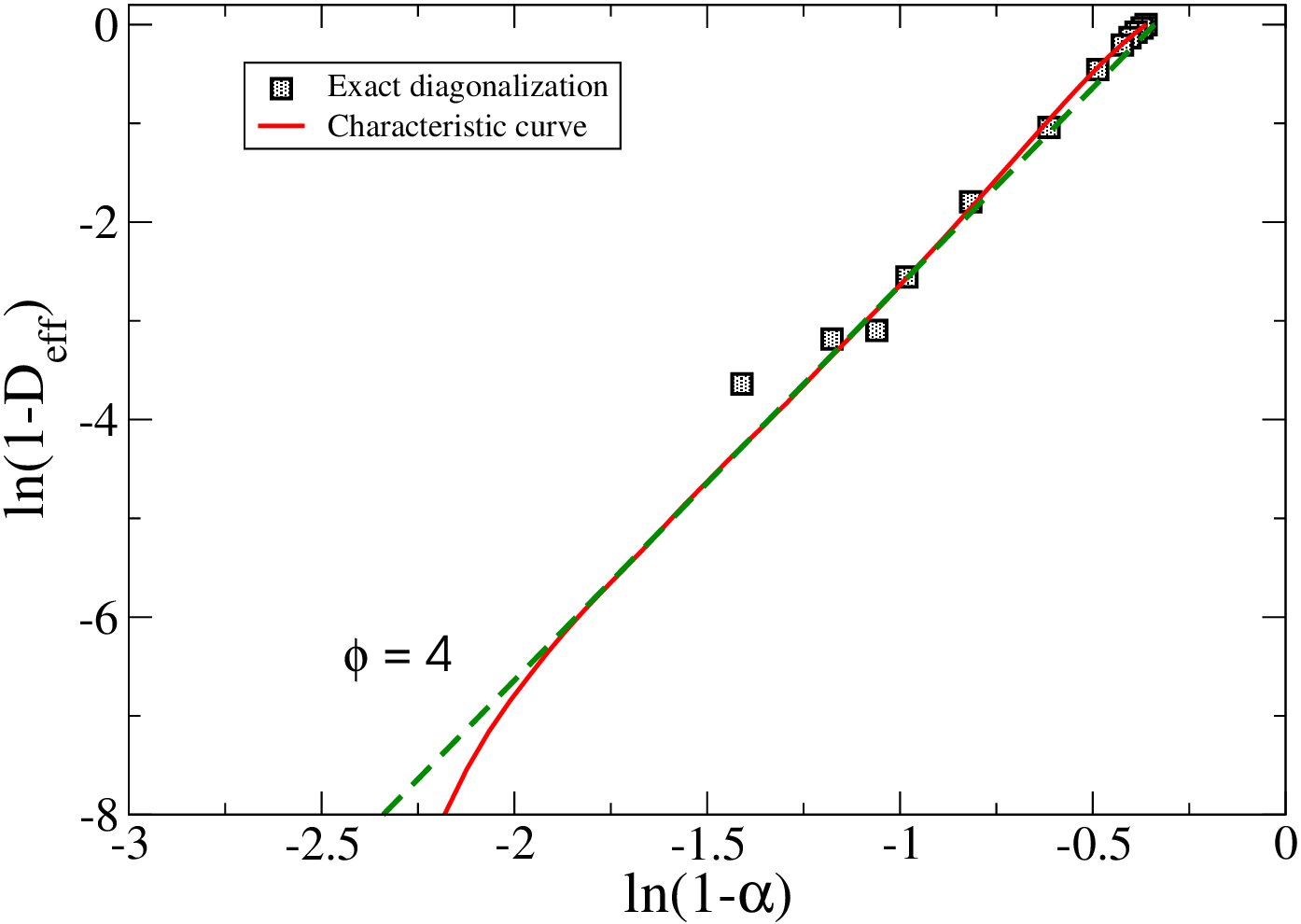}\\
\vspace*{0.2cm}
\caption{Comparison between a 'average' characteristic curve (solid line) 
and the exact diagonalization result (squares).
At the right panel we see that $\phi=4$ is a good estimative for
the crossover exponent.}
\label{lcext}
\end{center}
\end{figure}

\begin{figure}[h!]
\begin{center}
\vspace*{0.9cm}
\includegraphics[scale=0.4]{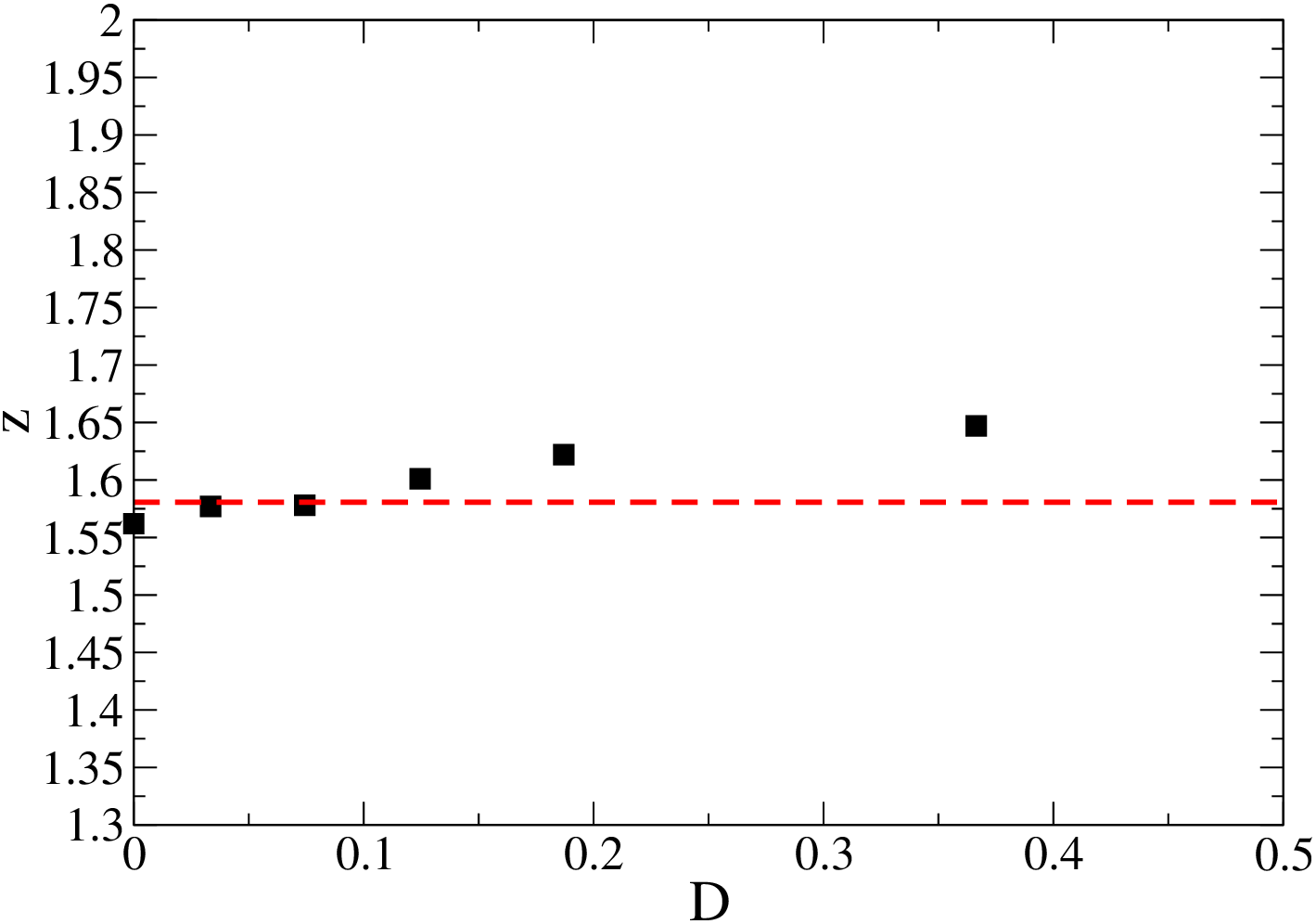}
\vspace*{0.2cm}
\caption{Estimative for the $z$ exponent calculated by some values
of the diffusion rate. The dashed line indicates the DP value.}
\label{zexp}
\end{center}
\end{figure}

On the other hand, a scaling behavior for $\Gamma$, with $D$ held fixed, is given by the
expression

\begin{eqnarray}
\Gamma=L^{-z}f(\Delta L^{1/\nu_{\perp}}),
\end{eqnarray}
where $\Delta=|\alpha_c-\alpha|$.
Defining the quantity

\begin{eqnarray}
Y_L(\alpha,D)=\frac{\ln[\Gamma(\alpha,D;L+1)/\Gamma(\alpha,D;L-1)]}{\ln[(L+1)/(L-1)]},
\end{eqnarray}
we may estimate the critical point $\alpha_c(L)$ finding the intersection
of the curves $Y_L$ and $Y_{L+1}$ \cite{carlon}. This procedure resembles
the phenomenological renormalization group \cite{plasc}. 
The sequence $\alpha_c (L)$
of estimates for a given value of $D$, with $L$ ranging between 4 and 14, 
was extrapolated to the thermodynamic limit, producing a critical line
shown in the figure \ref{lcext}. Although in the higher diffusion region
the coincidence between the two lines, shown in the figure, reduces 
there seems to be no doubt that
$\phi=4$ is a good estimate for the crossover exponent. Actually, 
the estimative for this exponent using the exact diagonalization
approach is $\phi=3.87\pm 0.17$. Probably for 
high diffusions is necessary to study larger lattices to determine
more precise results for the critical point as well as its exponent. 
The exponent $z$ is shown as function of the diffusion rate in the
figure \ref{zexp}. For smaller diffusion rate, the exponent value coincide
with that predicted by the DP universality class. However, for $D> 0.2$
occurs a growing of this estimative and a simple quadratic extrapolation   
gives $z=1.93$, near to multicritical point, which is in reasonable
agreement with mean-field value for this exponent, $z=2$. 

\section{Conclusion}

We study the difussive contact process in one-dimension analyzing 
its critical properties. A special attention is devoted for the
determination of the crossover between a classical mean-field 
behavior in the infinite diffusion limit and DP universality
class presents when the diffusion rate is finite.
To accomplish this task we use supercritical pertubative
series and exact solutions for finite-size lattices.

Calculating a supercritical series for the ultimate survival probability and 
analyzing it using PDA's we obtain estimates for the critical line 
of the CP model with diffusion.  The critical line was derived through 
the integration of the equation (\ref{pda}) by the method 
of the characteristics.  Direct results for the value of the crossover 
exponent using the scaling form \ref{mcs} using Styer's algorithm  
\cite{s90} we were not be able to be obtain with an acceptable precision.  
However the method of the characteristics permitted us to calculate
the critical 
line and, consequently, the value for the crossover exponent $\phi$.  
Our result, $\phi=4.02\pm0.13$ is in agreement with that derived 
in \cite{mariofiore} and explores a region of diffusion very close 
to the multicritical point.

The technique of the two-variable supercritical series associated 
with PDA analysis was shown to be accurate enough to determine the 
critical properties in similar models \cite{wgdstilckcross}.  
Therefore, we believe that a natural extension for this work 
is analyze related models that apparently possess non-trivial 
multicritical points. This seems to be the cases of the pair-creation
and triplet-creation  
models with diffusion, also studied in \cite{mariofiore}.  
This research is already in course.  

Finally, the results provided by the exact diagonalization
of the time evolution operator are according to simulational
data as well as that ones obtained by the series analysis. Actually,
in higher diffusion region, a greater fluctuation take place, suggesting
that this system must be studied in larger sizes. Nevertheless, the critical
curve indicates that $\phi\approx 4$ is a good estimate for the crossover
exponent.


\section*{Acknowledgement}
W.G. Dantas acknowledges the financial 
support from  Funda\c{c}\~ao de Amparo \`a Pesquisa do
Estado de S\~ao Paulo (FAPESP) under Grant No. 05/04459-1 and
JFS acknowledges funding by project  PRONEX-CNPq-FAPERJ/171.168-2003.

\end{document}